\documentstyle[epsf]{aa}
\newcommand{\ms}{M$_{\odot}$}

\newcommand{\teff}{$T_{\rm eff}$}
\newcommand{\lgg}{log\,$g$}
\newcommand{\kms}{km\,s$^{-1}$}

\newcommand{\ea}{et~al.~}

\newcommand{\mua}{$\mu_{\alpha}{\rm cos}\delta$}

\newcommand{\mud}{$\mu_{\delta}$}

\begin{document}
 
\title{Kinematical trends among the field horizontal branch 
stars\thanks{Based in part on HIPPARCOS data}}

\author{Martin Altmann\inst{}
\and Klaas S. de Boer\inst{}
}

\institute{Sternwarte, Univ. Bonn, Auf dem H\"ugel 71, D-53121 Bonn, Germany
}

\date{Received: 30.08.1999 / Accepted: 29.10.1999}

\thesaurus{04(08.08.2; 08.11.1; 08.16.3; 10.08.1; 10.11.1; 10.19.3)}

\offprints{maltmann@astro.uni-bonn.de}

\maketitle

\markboth{M. Altmann \& K.S. de Boer: Kinematics of horizontal branch stars}
        {M. Altmann \& K.S. de Boer: Kinematics of horizontal branch stars}

\begin{abstract}
Horizontal branch (HB) stars in the field of the Milky Way can be used 
as tracers for the study of early stages of the evolution of our galaxy.
Since the age of individual HB stars is not known a priori, 
we have studied the kinematics of a sample of field HB stars 
measured with Hipparcos to look for signs of age and population nature. 
Our sample comprises 14 HBA, 2 HBB and 5 sdB/O stars.
We found that the kinematics of the HBA stars is very different from 
that of the sdB/O stars (including those from an earlier study). 
The HBA stars have low orbital velocities, some are even on retrograde orbits. 
Their orbits have large eccentricities and in many cases reach
large distances above the galactic plane.
In contrast, the sdB/O stars show disk-like orbital characteristics.
The few HBB stars (with $T_{\rm eff}> 10,000$ K) in our sample seem to
have kinematics similar to that of the sdB/O stars. 

In order to see if there is a trend among the HB stars in their kinematics, 
we investigated also RR Lyrae stars measured with Hipparcos. 
Here we found a mixed kinematical behaviour, 
which was already known from previous studies. 
Some RR Lyrae stars have disk-like orbits (most of these being metal rich) 
but the majority has halo-like orbits, 
very similar to those of our HBA stars.

Since the atmospheres of most types of HB stars do not reflect original 
metallicities any more 
the kinematics is the only aspect left to study the 
origin and population membership of these stars. 
Thus, the clear trend found in kinematics of stars along the HB,
which is also a sequence in stellar mass, 
shows that the different kinds of field HB stars arose from stars 
having different origins in age and, e.g., metallicity or mass loss rate.

\keywords{Stars: kinematics -- Stars: horizontal branch -- Stars: Population II
 -- Galaxy: halo -- Galaxy: kinematics and dynamics -- Galaxy: structure} 
\end{abstract}

\section{Introduction: HB-stars, their population membership and the galactic structure}

Stars of horizontal-branch nature are important objects 
in studies of the older stellar populations 
and in studies of galactic structure in relation with 
formation theories for the Milky Way. 
Their virtue lies in three properties. 
First, they have rather well defined absolute magnitudes 
and their distances are therefore easy to determine.  
Second, their nature is such that they are easy to discover, 
in particular at higher galactic latitudes. 
Third, the stellar evolution leading to such stars is 
in principle 
relatively well understood. 

The horizontal branch (HB) stellar phase is the 
core helium burning late stage of evolution of originally 
low to medium mass stars. 
During the red giant phase the stars have lost mass such 
that a helium core of $\simeq 0.5$~\ms, 
surrounded by an outer shell of hydrogen gas, remains. 
For very thin shells, $M_{\rm shell} < 0.02$ \ms, 
the stars are rather blue and of spectral class sdB (subdwarf B), 
for ever thicker shells their atmospheres are cooler, 
leading to spectral types such as horizontal branch B and A 
(HBB and HBA) and to the red HB (RHB) stars 
with $M_{\rm shell} \simeq 0.5$ \ms. 
Between the HBA and RHB stars lies the 
pulsational instability strip with the RR Lyrae stars.  
(For the systematics of the spectral classification see 
de Boer et al. 1997c.) 
HBA and HBB stars are often called blue HB (BHB) stars, 
the even hotter sdB and sdO stars are also known as extended 
or extreme HB (EHB) stars.

BHB and EHB stars can be easily found based on their blueness, 
the RR Lyr stars due to their variability.
Of the BHB stars, the HBAs are easily identified, 
as their physical parameters differ from main sequence A stars, 
while HBBs may be confused with main sequence B stars. 
However, at higher galactical latitudes A or B main sequence stars 
are very rare.
RHB stars can easily be confused with subgiants and giants 
because they lie in the same region of the HRD.

To trace the structure of the (local part of the) Milky Way
one employs several techniques (see also the review by Majewski 1993). 
The classical method is to perform star counts 
(see, e.g., Bahcall \& Soneira 1984).
Selecting stars based on their proper motion allows, 
if their radial velocities and distances are also determined, 
to study the true kinematics of the stars 
(see, e.g., Carney \ea 1996 and papers cited there). 
Basing oneself on proper motions one naturally studies the general group 
of more nearby stars extended by true high-velocity stars. 
Another method is to sample distances and radial velocities of 
a specific set of stars, such as BHB or RR Lyr stars, 
to investigate these parameters in a statistical manner 
(see, e.g., Kinman \ea 1994, 1996). 
A more specific method is to observe statistically complete samples 
of stars of a special type in several directions to derive scale heights 
(for sdB stars see e.g., Heber 1986, Moehler \ea 1990, Theissen \ea 1993), 
or scale lengths in the Milky Way.
A further possibility is to go beyond the present kinematic parameters 
by calculating orbits 
based on distances, radial velocities, and proper motions. 
This method has been used for high proper motion stars (Carney \ea\ 1996),
other dwarf stars (Schuster \& Allen 1997), 
and for globular clusters (Dauphole \ea\ 1996). 
Also the orbits of sdB stars have been investigated 
(Colin \ea\ 1994, de Boer \ea\ 1997a). 
The latter study showed that most of the sdB stars have orbits 
staying fairly well inside the Milky Way disk, 
indicating that the sdB stars are not generally part of the halo population. 

In the present study we have attempted to perform a similar 
analysis for HBA and HBB stars (for short: HBA/B stars).
Our sample consists of the local HB stars 
which were observed by the Hipparcos satellite. 
These are the HB-like stars with the most accurate spatial and kinematic data 
available to date. 

However, only for a few HBA/B stars are the parallaxes accurate enough 
to calculate reliable distances (de Boer \ea\ 1997b). 
For the other stars the distance still must be derived from photometry. 
Here one needs to know the absolute magnitude of field horizontal branch stars.
Especially since the publication of the Hipparcos catalogue
a lot of effort has gone into fixing this value. 
However, this has not yet led to total agreement. 
For a review of various approaches to solving this problem we refer to 
de Boer (1999) and Popowski \& Gould (1999).

An important parameter in these studies is the metallicity of the stars, 
as it is generally thought to be correlated with age. 
For dwarf stars metallicities can be estimated using 
photometric indices or spectroscopy (see the summary by Majewski 1993). 
For HB stars this is, unfortunately, not a trustworthy method. 
The atmospheres of many HB stars have most probably been altered 
chemically with respect to the original composition. 
Gravitational settling of heavy elements in the sdB/O star 
and possibly HBB star atmospheres
leads to a present lower content of elements like He,
while levitation of heavy elements 
leads to atmospheres with enhanced abundances of certain elements like Fe or Au
as found in several field horizontal branch stars, 
e.g. Feige 86 (Bonifacio et al. 1995).
Levitation must also be the explanation for the high metal abundances 
in blue HB stars in M 13 (Behr \ea\ 1999) and NGC 6752 (Moehler \ea\ 1999) 
finally uncovered to explain deviant flux distributions near the 
Balmer jump of globular cluster blue HB stars (Grundahl \ea\ 1999). 
Therefore, original metallicities (as well as the original masses) 
are no longer accessible quantities. 
Determining the kinematic properties can help deciding
which  of the HB stars are intrinsically more metal poor and which are more 
metal rich, and hence of somewhat younger origin.

The main subjects of our study are the HBA/B stars 
which are located in the colour magnitude diagram on the horizontal branch
between the RR-Lyrae stars and the hot subdwarfs. 
Unfortunately the main sequence crosses the HB at the HBB region, 
so that HBB stars can be confused with normal B stars. 
Therefore we have only few HBB stars in our sample. 
We mainly focus on HB stars with temperatures lower than 10000 K
which lie above the main sequence.

Sect. 2 deals with the data neccessary for our study.
 In Sect. 2.3 we determine the absolute magnitudes and distances of 
the HBA/B and sdB/O  stars with the method of {\it auto-calibration} using the shape 
of the HB defined by the stars with the best Hipparcos parallaxes. 
 In Sect. 3 we discuss the kinematical behaviour of the HBA/B and sdB/O stars and 
make comparisons with the results of de Boer \ea\ (1997a). 
 To further explore a possible trend in kinematics of stars along the HB 
we investigate (Sect. 4) the orbits of a sample of RR-Lyrae stars.

\begin{table*}
\caption[]{Physical properties of the sample of horizontal branch stars.}
\begin{tabular}{lrrrrrlrrl}
\hline
\multicolumn{1}{c}{Name}
& \multicolumn{1}{c}{HIP}
& \multicolumn{1}{c}{$V^{a}$}
& \multicolumn{1}{c}{$B-V^{a}$}
& \multicolumn{1}{c}{$E_{B-V}$$^{b}$}
& \multicolumn{1}{c}{$\delta {M_V}^{b}$}
& \multicolumn{1}{c}{Type$^{c}$}
& \multicolumn{1}{c}{\teff}
& \multicolumn{1}{c}{\lgg}
& \multicolumn{1}{c}{Source$^{d}$}\\
& &
 \multicolumn{1}{c}{[mag]} & \multicolumn{1}{c}{[mag]}
& \multicolumn{1}{c}{[mag]}
&\multicolumn{1}{c}{[mag]} & & \multicolumn{1}{c}{[K]} & & \\
\hline
 HD~2857       &   2515  &  9.967 &   0.219 & 0.050 & $-$0.001 & HBA     &  7700 & 3.1  & GCP, HBC\\
 HD~14829      &  11124  & 10.228 &   0.023 & 0.020 & $-$0.580 & HBA     &  8700 & 3.3  & GCP\\
 HD~60778      &  36989  &  9.131 &   0.135 & 0.020 & $-$0.040 & HBA     &  8600 & 3.3  & GCP, S91, HBC\\
 HD~74721      &  43018  &  8.717 &   0.042 & 0.000 & $-$0.330& HBA     &  8600 & 3.3  & GCP, S91, HBC\\
 HD~78913      &  44734  &  9.291 &   0.094 & 0.030 &$-$0.215& HBA     &  8700 & 2.5  & IUE-fit, S91 \\
 HD~86986      &  49198  &  8.000 &   0.119 & 0.035 &$-$0.130 & HBA     &  7900 & 3.1  & B97b, S91\\
 BD~+36 2242   &  59252  &  9.904 &$-$0.065 & 0.010 &$-$1.188& HBB     & 11400 & 4.4  & HBC \\
 HD~106304     &  59644  &  9.077 &   0.027 & 0.040 &$-$0.696& HBA     &  9500 & 3.0  & IUE-fit, S91\\
 BD~+42 2309   &  60854  & 10.820 &   0.043 & 0.000 &$-$0.324 & HBA     &  8400 & 3.3  & GCP\\
 HD~109995     &  61696  &  7.603 &   0.047 & 0.001 & $-$0.307 & HBA     &  8300 & 3.15 & B97b, S91  \\
 BD~+25 2602   &  64196  & 10.148 &   0.057 & 0.065 & $-$0.659 & HBA     &       &      & S91\\
 HD~117880     &  66141  &  9.059 &   0.082 & 0.015 & $-$0.201 & HBA     &  9200 & 3.4  & GCP, S91, HBC\\
 Feige 86      &  66541  & 10.006 &$-$0.140 & 0.050 & $-$2.193 & HBB     & 15300 & 4.1  & HBC \\
 HD~130095     &  72278  &  8.155 &   0.032 & 0.064 &$-$0.840 & HBA     &  8800 & 3.15 & B97b, S91\\
 HD~139961     &  76961  &  8.857 &   0.098 & 0.107 &$-$0.666 & HBA     &  8750 & 3.3  & B97b, S91\\
 HD~161817     &  87001  &  7.002 &   0.166 & 0.020 &$-$0.001 & HBA     &  7500 & 2.95 & B97b, S91\\
\hline
 CD~$-$38 222  &   3381  & 10.400 &$-$0.224 & 0.013 & $-$2.639 & sdB     & 28200 & 5.5  & B97b, B97a\\
 Feige 66      &  61602  & 10.602 &$-$0.286 & 0.040 & $-$3.556 & sdB     & 28000 & 4.9  & KHD, S94\\
 HD~127493     &  71096  & 10.040 &$-$0.251 & 0.095 & $-$3.756 & sdO     & 40000 & 5.8  & KHD\\
 HD~149382     &  81145  &  8.872 &$-$0.280 & 0.050 & $-$3.598 & sdOB    & 40000 & 5.8  & KHD, S94\\
 HD~205805     & 106917  & 10.158 &$-$0.241 & 0.025 & $-$2.929 & sdB     & 25000 & 5.0  & B97b, B97a\\
\hline
\end{tabular}\\
$^{a)}$ $V$, $B-V$ from the Hipparcos Catalogue\\
$^{b)}$ $E_{B-V}$, $\delta M$, see Sect. 2.2 and 2.3\\
$^{c)}$  Type:
      HBA stars: \teff\ $<$ 10500 K;
      HBB stars: 20000 K $>$ \teff\ $\ge$ 10500 K;
      sdB/O stars: from literature (see under $source$)\\
$^{d)}$ The values for \lgg\ and \teff\ have been taken from the first work cited.
HBC: Huenemoerder et al. (1984),
GCP: Gray et al. (1996),
B97a: de Boer et al. (1997a),
B97b: de Boer et al. (1997b),
KHD: Kilkenny et al. (1987) and references therein,
S91: Stetson (1991),
S94: Saffer et al. (1994),
IUE-fit: see Sect. 2.2
\end{table*}

\begin{table*}
\caption[]{Spatial and kinematical data for the stars$^a$ of our sample}
\begin{tabular}{lllrrrrrrrrl}
\hline
Name  & \multicolumn{2}{l}{RA (Eq. 2000.0) DEC } & 
        \mua & \mud & $\Delta$\mua & $\Delta$\mud & $\pi$ & $\Delta\pi$ & $d$ & {$v_{\rm rad}$}
        & ref.$^b$\\
        & hr, min, sec & \ \ \ \ $^\circ$ \ \ $\arcmin$ \ \ $\arcsec$ & 
        mas/yr &  mas/yr &  mas/yr & mas/yr & mas & mas & pc & km\,s$^{-1}$ & {$v_{\rm rad}$}\\
\hline 
 HD~2857    &   00 31 53.80 & $-$05 15 42.3  &  $-$6.85 & $-$66.05 & 1.58 
& 0.85 & 1.79 & 1.67 & 687 & $-149$ & CG \\
 HD~14829   &  02 23 09.23 & $-$10 40 38.9  & +31.31 & $-$46.85 & 1.89 & 1.58 & 4.40
& 1.96 & 619 & $-176$ &  P69 \\
 HD~60778   &  07 36 11.79 & $-$00 08 14.9  & $-$20.92 & $-$84.04 & 1.21 & 0.73 & 2.35
& 1.23 & 479 &  $+39$ & eE \\ 
 HD~74721   &  08 45 59.29 & +13 15 49.6  & $-$41.83 &$-$112.42 & 1.31 & 0.93 & 0.34
& 1.46 & 356 &   +9 &  eE \\ 
 HD~78913   &  09 06 54.78 & $-$68 29 22.1  & +36.48 &  $+$22.87 & 0.95 & 0.80 & 2.69
& 0.88 & 469 & $+313$ & cE \\
 HD~86986   &  10 02 29.48 & +14 33 27.0  & +144.06 &$-$208.27 & 1.05 & 0.53 & 3.78
& 0.95 & 267 &   +9 & bE \\
 BD~+36 2242&  12 09 15.84 & +35 42 42.9  &  $-$5.57 &  $-$1.83 & 1.20 & 0.88 & 1.83
& 1.25 & 409 &   $-$4 &  dE \\ 
 HD~106304  &  12 13 53.63 & $-$40 52 23.7  & $-$90.36 &$-$117.00 & 0.99 & 0.71 & 2.83
& 1.12 & 336 &  +95 &  cE \\ 
 BD~+42 2309&  12 28 22.18 & +41 38 52.7  & $-$21.30 & $-$33.61 & 1.19 & 1.39 & 0.47
& 1.76 & 942 & $-152$   &  dE \\ 
 HD~109995  &  12 38 47.69 & +39 18 32.9  &$-$114.81 &$-$144.19 & 0.83 & 0.68 & 4.92
& 0.89 & 215 &  $-132$   & BB \\
 BD~+25 2602&  13 09 25.64 & +24 19 25.3  & $-$84.51 & $-$18.73 & 1.83 & 1.44 & 1.40
& 1.54 & 540 &  $-$74 &  eE \\ 
 HD~117880  &  13 33 29.86 & $-$18 30 53.1  & $-$85.65 &$-$140.33 & 1.18 & 0.75 & 4.80
& 1.10 & 433 &  $-$45 & cE \\
 Feige 86   &  13 38 24.77 & +29 21 57.0  & $-$15.34 &$-$109.79 & 1.49 & 0.92 & 4.61
& 1.65 & 255 &  $-$22 &  cE  \\
 HD~130095  &  14 46 51.35 & $-$27 14 53.3  &$-$213.89 & $-$79.77 & 1.29 & 0.75 & 5.91
& 1.08 & 199 &  $+58$  & BB \\
 HD~139961  &  15 42 52.97 & $-$44 56 40.0  &$-$187.04 & $-$92.41 & 1.28 & 1.19 & 4.50
& 1.19 & 280 & +145 &  dE \\
 HD~161817   & 17 46 40.65 & +25 44 57.3  & $-$37.05 & $-$43.23 & 0.50 & 0.57 & 5.81
& 0.65 & 183
 & $-$363 & bW \\
\hline
CD~$-$38 222 &   00 42 58.28 & $-$38 07 37.2  &  $+$43.85 &  $-$7.00 & 1.88 & 1.23 & 3.07
 & 1.73 & 262 &  $-$52 & GS \\
 Feige 66   &  12 37 23.52 & +25 04 00.1  &  $-$2.72 & $-$26.71 & 1.80 & 1.36 & 5.11
& 1.74 & 182 &   +1 &   cE \\
 HD~127493  &  14 32 21.51 & $-$22 39 25.5  & $-$32.80 & $-$17.22 & 1.45 & 1.37 & 5.21
& 1.49 & 118 &  +13 &   bW \\
 HD~149382  &  16 34 23.34 &  $-$04 00 52.0  &  $-$5.95 &  $-$3.92 & 1.83 & 1.73 &13.07
& 1.29 &  79 &   +3 &   cW \\
 HD~205805   & 21 39 10.55 & $-$46 05 51.4  &  $+$76.39 &  $-$9.93 & 1.20 & 0.90 & 3.77
& 1.70 & 201 &  $-57$   & B97a  \\
\hline
\end{tabular}\\
$^{a)}$
Positions, proper motions and parallaxes (with errors) listed in this table are from the Hipparcos Catalogue,
the distances, as derived in Sect. 2.3.\\
$^{b)}$
References for radial velocities:
E: The Revision of the General Catalogue of Radial Velocities (Evans 1967),
W: The General Catalogue of Radial Velocities (Wilson 1953);
here the small case letters indicate the quality of the radial velocity:\\
a: $\Delta v_{\rm rad}<$ 0.9 \kms, b:$\Delta v_{\rm rad}<$ 2.0 \kms, 
c: $\Delta v_{\rm rad}<$ 5.0 \kms,
d: $\Delta v_{\rm rad}<$ 10.0 \kms, e: $\Delta v_{\rm rad}>$ 10.0 \kms.\\
B97a: de Boer \ea\ (1997a),
BB: Barbier-Brossat (1989),
CG: Corbally \& Gray (1996), 
GS: Graham \& Slettebak (1973),
P69: Philip (1969)
\end{table*}

\section{The data}
\subsection{Composition of the sample}

Our sample consists of the Hipparcos (ESA 1997) measured HB stars. 
In order to identify them we searched through lists of  bright HB-candidates 
in publications concerning horizontal branch stars, such as Kilkenny 
\ea (1987) for sdB/O stars, and Corbally \& Gray (1996), Huenemoerder 
\ea (1984), and de Boer \ea (1997b) for the HBA/B stars. 
However, for a few stars in these lists indications exist
that they are probably not horizontal branch stars.
 Among these are HD~64488 (Gray et al. 1996), 
HD~4772 (Abt \& Morell 1995, Philip et al. 1990),
HD~24000 (Rydgren 1971), HD~52057 (Waelkens et al. 1998, Stetson 1991) 
and HD~85504 (Martinet 1970). This sample, although being of limited size, 
represents the HB stars with by far the best kinematical data 
currently available.

Two further stars are HB-like but were excluded from the study nevertheless. 
BD~+32 2188 has a rather low value for  
 \lgg\ so that it lies considerably above the ZAHB
in the \lgg\ $-$ \teff\ diagram.
Being metal deficient (Corbally \& Gray 1996) it can be considered a
 horizontal branch star evolving away from the ZAHB. 
Because the evolutionary state is not fully HB 
the star cannot be part of our sample.
HD~49798 is a subluminous O-star. However, its \lgg\ is relatively low and
its trigonometrical parallax implies a star with absolute brightness
of about $-$2 mag, far too bright for a normal sdO star. 
It is probably on its way from the horizontal branch to become a 
white dwarf or it is a former pAGB star.
Because of these aspects we excluded this star. 

A large fraction of the known horizontal branch stars has no published radial velocity and could
therefore not be used for our study.
A few stars had radial velocities but no Hipparcos data.

There is no constraint on the position, so that the sample stars are located in all parts of the sky.
However, as many studies were made in fields near the galactic poles we have relatively more
stars at very high galactic latitudes.
 Although our sample of stars is certainly not
statistically complete in any way, we do not expect noticeable selection effects
due to position in the sky (see Sect. 5.2.).

\subsection{Physical properties of the stars, extinction}
While many of the stars are classical template HB stars, like HD~2857,
HD~109995, HD~130095 or HD~161817, others 
are not as well studied.

For most of our stars values for $\log g$ and $T_{\rm eff}$ 
are available in the literature from a variety of methods. 
Sources are given in Table 1. 
For HD 78913 and HD 106304 $\log g$ and $T_{\rm eff}$ were derived 
from a fit of Kurucz models to spectrophotometric IUE data and photometry. 
For BD~+25~2602 no data are available to determine $\log g$ and $T_{\rm eff}$.
We keep it as part of our sample, as it was 
identified as a horizontal branch star by Stetson (1991).

Wherever possible we took the values for $E_{B-V}$ from de Boer \ea\ (1997b), 
supplemented by values listed in Gratton (1998).
For the other stars we derived the $E_{B-V}$, 
with ($B-V$)- and ($U-B$)-values taken from the SIMBAD archive
and a two-colour-diagram. 
Note that with this method there may well be metallicity dependent effects
having an influence on the reddening derived. 
For the star CD~$-$38~222 no ($U-B$) data are available; the reddening 
is very small as follows from the IRAS maps of Schlegel \ea\ (1998). 
We adopted the value from that study.

\subsection{Absolute magnitudes and distances}
We obtained the distances of the HB stars using the absolute magnitude of the relevant portion of 
the HB rather than directly using the Hipparcos parallaxes. The reason for this is that
most of the parallaxes are smaller than 3 mas which means that their
error of on average 1 mas is too large to calculate accurate distances. The absolute magnitudes $M_V$,
which are a function of the temperature and thus of $(B-V)_0$, have been derived through
self calibration as follows.

We started with the determination of the shape of the field horizontal branch.
For this we calculated the absolute magnitudes of those HB and sdB/O stars which have 
reasonably good parallaxes. 
For the determination of the mean absolute magnitude of the HB sample we excluded HD~74721 and BD~+42~2309
because their absolute magnitudes, calculated from their parallaxes, are too bright by more than 3.5 magnitudes.
Also excluded at this point are HD~14829 and HD~117880, whose parallaxes lead to absolute magnitudes far too faint.
With this medianization (leaving out the extremes to both sides) we ensure that our result is not affected
by stars with extreme values. Furthermore the stars having parallaxes with $\Delta\pi/\pi>1$ were excluded for the 
determination of the shape of the HB.

We then fitted by eye a curve to our sample in the colour magnitude diagram. In order to smooth this curve, it was approximated
by a polynomial.
Note that we aim to fit the observed parameters of the field horizontal branch and that we do not rely on a shape taken from
globular clusters or theoretical models (see Fig. 1).

From this we determined the value $\delta M_V$ giving the difference of $M_V$ for each ($B-V$)$_0$
with respect to $M_V$ at ($B-V$)$_0$=0.2 mag. Although the available metallicity measurements show a large spread 
for individual stars (see table II of Philip 1987),
 the averages for each lie around [Fe/H] $\sim-$1.5 dex.
Since the effect of metallicity on $M_V$ is small for RR Lyr stars (about 0.1 mag per 0.5 dex, see de Boer 1999)
we will neglect the metallicity effects for the HBA stars.

 Distances and absolute magnitudes of a sample of stars
obtained through trigonometric parallaxes
have to be corrected for the Lutz-Kelker bias (Lutz \&\ Kelker 1973).
This statistical effect, depending on the relative error of the parallaxes,
leads to an over-estimation of the
parallax on average, leading to too faint absolute magnitudes and
too short distances of the sample. 

The correction we applied is based on the averaging of parallaxes.
For that we have to correct the parallaxes of individual stars, acknowledging
that such a correction is only valid in a statistical sense.
The expected parallax $\pi^*$ given by
\begin{equation}
\pi^*=10^{0.2\left[ M_V-V-\delta M_V\right]-1+0.62E_{B-V}}
\end{equation}
with $M_V$ being the absolute magnitude, $V$ the apparent magnitude, $E_{B-V}$
the reddening. $\delta M_V$ is a term which accounts for the temperature and/or
$B-V$ dependence of the absolute magnitude of BHB stars in the same way as done by Gratton (1998).
Now $M_V$ is varied and \\
\mbox{$\chi^2$($M_V$)=$\sum_i(\pi^*_i(M_V)-\pi_i)^2/{(\Delta\pi_i)}^2$} is
 calculated (formula as revised by Popowski \& Gould 1999).
 At the correct $M_V$ the average of $\chi^2$ should be minimized.

\begin{figure}
\def\epsfsize#1#2{1.0\hsize}
\centerline{\epsffile{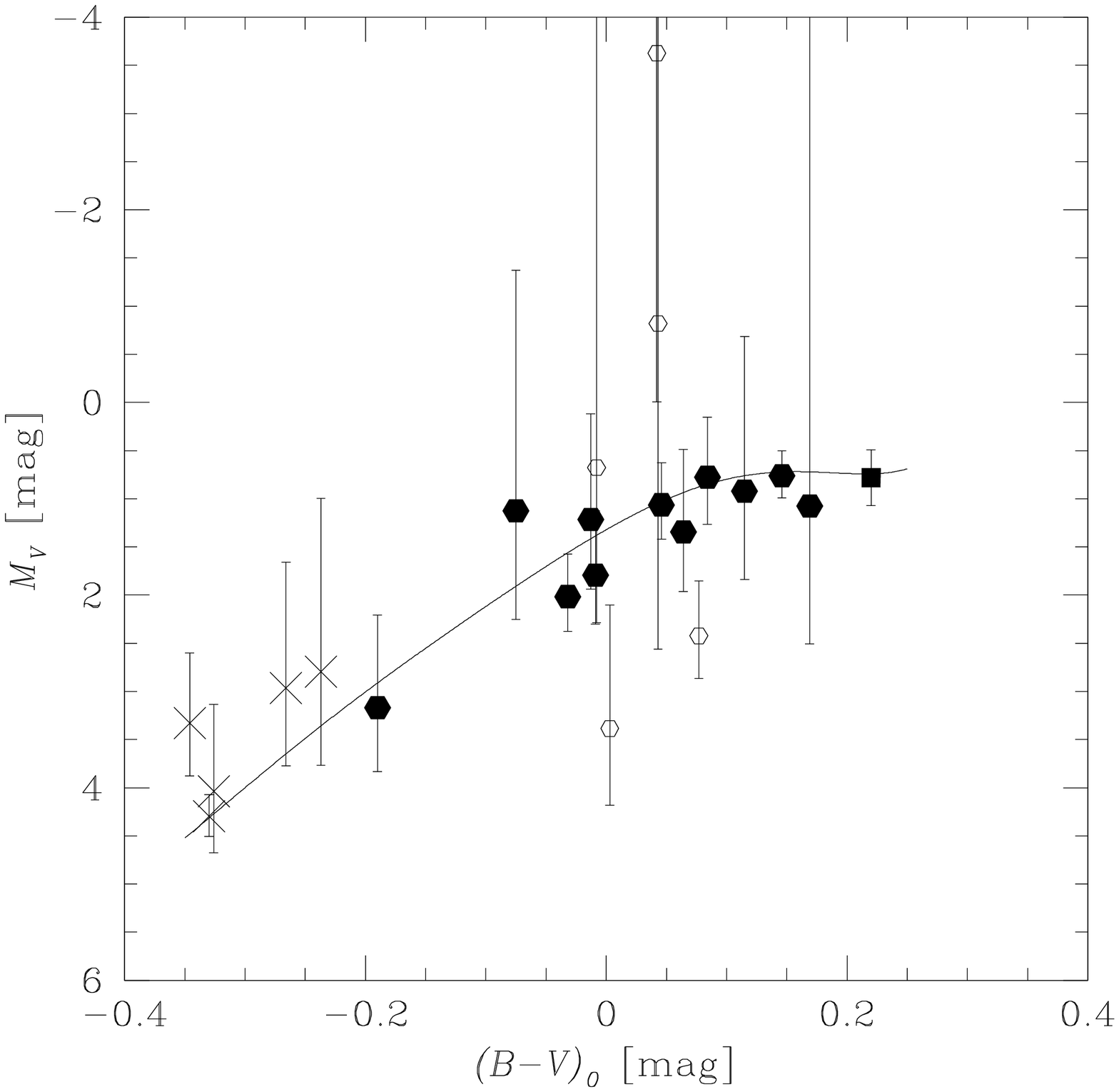}}
\caption[]{Colour magnitude diagram showing the stars of our sample and the curve
which was used as the shape of the FHB.\\
Hexagons: HBA/B stars (open symbols mean stars not used for the fit),
crosses mean sdB/O stars, the square depicts RR Lyrae.
}
\end{figure}

$M_V$ is now found using all stars, regardless of their $\Delta\pi/\pi$, except the four excluded above.
We arrived at an absolute magnitude of $M_V=0.63\pm 0.08$ mag for the horizontal part ($(B-V)_0\sim 0.2$ mag) 
of the horizontal branch. As stated before this value 
should be valid for [Fe/H]$\sim -$1.5 dex.
However, as the curve defining the shape of the HB is subjective to a certain extent
the real error of the HB's absolute magnitude is somewhat larger. 
The absolute magnitudes and thus the distances of the individual stars including those omitted earlier
are obtained by adding their $\delta M_V$ to the mean absolute magnitude of the HB. 

\subsection{Proper motions and positions}
Positions and proper motions used in this work were taken from the 
Hipparcos catalogue (ESA 1997). 
The mean error of the proper motions is below 1.5 mas/yr
(see Table 2) which means an error in the tangential velocity of 3.5 \kms\
for a star at a distance of 500 pc. 
As most of our stars have smaller distances 
the error caused by the proper motion uncertainty is even smaller. 

No star of the sample of HBA/B or sdB/O stars has an astrometric flag in the Hipparcos catalogue, 
indicating there were no problems in the data reduction.
The Hipparcos goodness-of-fit statistic is below +3 in all cases,
meaning that the astrometric data derived from the Hipparcos catalogue should be reliable
and there are no indications that our sample contains double stars.

\begin{table*}
\caption[]{Orbital and kinematical characteristics}
        \label{tableofcoordinates}
\begin{tabular}{lrrcrrccrrrrrc}
\hline
Name &  {$R_{\rm a}$} & {$R_{\rm p}$} & {$z_{\rm max}$} 
 & {$ecc$} & {$nze$} & $U$ & $V$ & $W$ & {$\Theta$} & {$I_{\rm z}$} & {Type}\\
       & {[kpc]} & {[kpc]} & {[kpc]} & & & {[km\,s$^{-1}$]} & 
{[km\,s$^{-1}$]} & {[km\,s$^{-1}$]} & {[km\,s$^{-1}$]} & {[kpc\,km\,s$^{-1}$]} & \\
\hline
 HD~2857     & 11.81 & 0.44 &  6.47 & 0.93 & 1.04 &  +156 &  +25 &   +67 &  $+$29 &  $+$251 & HBA \\
 HD~14829    & 11.29 & 2.69 &  8.11 & 0.62 & 1.01 &  +108 &  +71 &  +156 &  $+$71 &  $+$622 & HBA \\
 HD~60778    & 9.35  & 2.41 &  4.56 & 0.59 & 0.56 &   +53 &  +82 &$-$115 &  $+$80 &  $+$714 & HBA \\
 HD~74721    & 8.65  & 1.95 &  4.42 & 0.63 & 0.59 &   +22 &  +70 &$-$109 &  $+$69 &  $+$606 & HBA \\
 HD~78913    & 9.52  & 1.79 &  0.38 & 0.68 & 0.05 &  +107 &$-$77 &   +24 &  $-$83 &  $-$695 & HBA \\
 HD~86986    & 16.96 & 0.33 & 13.26 & 0.96 & 1.63 &  +248 &  +24 &   +50 &  $+$20 &  $+$217 & HBA \\
 BD~+36 2242 & 9.96  & 8.58 &  0.42 & 0.04 & 0.05 &    +3 & +227 &    +3 & $+$227 & $+$1945 & HBB \\
 HD~106304   & 8.42  & 1.65 &  7.27 & 0.67 & 1.62 & $-$22 &  +38 &$-$150 &  $+$39 & $+$327  & HBA \\
 BD~+42 2309 &  9.96 & 0.91 &  5.26 & 0.83 & 0.89 &   +29 &  +34 &$-$110 &  $+$35 &  $+$300 & HBA \\
 HD~109995   &  9.41 & 0.48 &  5.52 & 0.90 & 1.00 &    +4 &  +30 & $-$96 &  $+$30 &  $+$258 & HBA \\
 BD~+25 2602 & 10.80 & 1.51 &  1.71 & 0.75 & 0.18 &$-$146 &  +72 & $-$53 &  $+$72 &  $+$607 & HBA \\
 HD~117880   &  8.91 & 4.17 &  9.29 & 0.36 & 4.89 & $-$55 &$-$28 &$-$199 &  $-$26 &  $-$217 & HBA \\
 Feige 86    &  9.18 & 2.95 &  0.27 & 0.51 & 0.03 &   +76 & +117 &  $-$7 & $+$118 & $+$995  & HBB \\
 HD~130095   &  8.94 & 0.49 &  5.13 & 0.90 & 0.93 & $-$58 &  +30 &   +65 &  $+$31 &  $+$258 & HBA \\
 HD~139961   &  8.26 & 1.57 &  1.68 & 0.68 & 0.22 &    +3 &$-$69 &   +81 &  $-$69 &  $-$568 & HBA \\
 HD~161817   & 12.51 & 1.30 &  7.36 & 0.81 & 0.74 &$-$169 &$-$54 &$-$129 &  $-$56 &  $-$473 & HBA \\
\hline
 CD~-38 222  &  9.31 & 7.29 &  1.23 & 0.12 & 0.13 & $-$38 & +206 &   +59 & $+$207 & $+$1749 & sdB \\
 Feige 66    & 9.10  & 7.76 &  0.22 & 0.08 & 0.02 &   +24 & +217 &    +8 & $+$217 & $+$1841 & sdB \\
 HD~127493   & 8.53  & 7.71 &  0.20 & 0.05 & 0.02 &   +10 & +212 &   +15 & $+$212 & $+$1780 & sdO \\
 HD~149382   & 9.64  & 8.33 &  0.13 & 0.07 & 0.01 &   +13 & +233 &   +10 & $+$233 & $+$1966 & sdB \\
 HD~205805   & 11.48 & 6.72 &  0.19 & 0.26 & 0.02 & $-$82 & +225 &  $-$3 & $+$225 & $+$1880 & sdB \\
\hline
\end{tabular}\\
Note: Due to a change in the convention (see Geffert 1998), the values of $I_z$ have changed 
their sign (positive $\Theta$ have now positive $I_{\rm z}$) 
in contrast to previous work (e.g. de Boer \ea\ 1997a)
\end{table*}

\begin{figure*}
\def\epsfsize#1#2{1.0\hsize}
\centerline{\epsffile{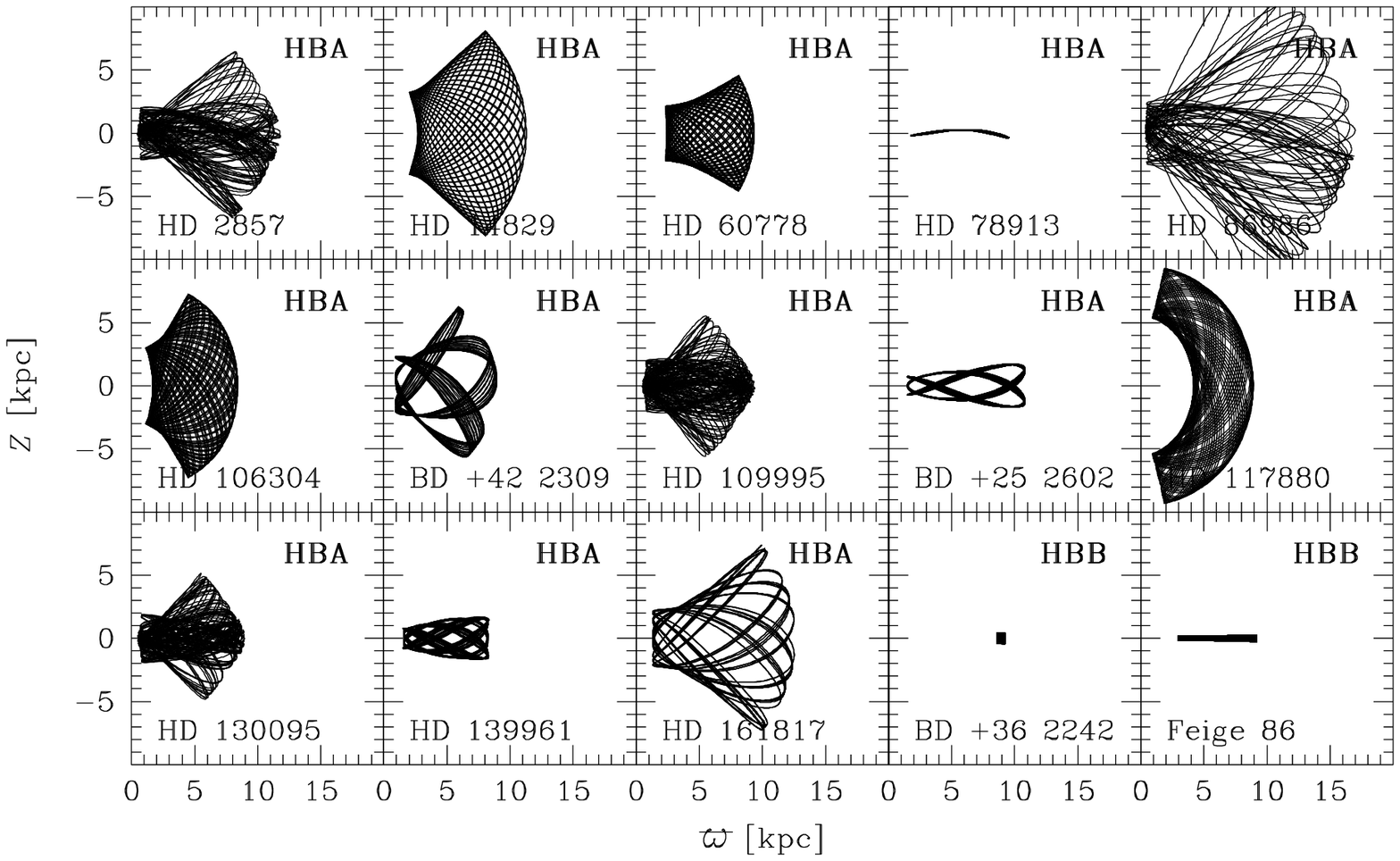}}
\caption[]{Orbits of the HBA/HBB stars displayed in meridional cuts.
The orbits shown here have been calculated for 10 Gyr, in order to make the shape
of the orbit better visible (The orbit of HD 74721 (similar to HD 60778) is not shown).
 For orbits of sdB stars see de Boer \ea\ (1997a).}
\end{figure*}

\subsection{Radial velocities}
The radial velocities were taken from original sources (see Table 2),
in part found from the Hipparcos Input Catalogue (Turon et al. 1987). 
The typical uncertainties are about 10 \kms, so that they should not
have a large effect on our results. 
The size of our sample was
limited to a large extent by the lack of radial velocities; for only about
30\% of the HB-candidates radial velocities could be found.

Radial velocities can be affected by binarity of the star. 
We cannot absolutely exclude this possibility for some of the stars, 
but as noted in Sect. 2.4 there are no indications for binary nature 
for any of our stars. 

For some stars, Corbally \& Gray (1996) found
drastically different values for the radial velocity. 
They note however that in many of these cases their values may 
be affected for some reason (see their Sect. 4) 
as they show strong deviations with respect to values from the literature. 
We therefore used radial velocities from Corbally \& Gray only
for HD~2857 for which no other value is available.

\section{Kinematics and orbits}
In order to gain information about the nature and population membership
of the stars we analyse their kinematic behaviour and calculate their
orbits.

\subsection{Calculating orbits and velocities}
Before calculating the orbits the observational data have to be transformed
into the coordinates of the galactic system ($X,Y,Z;U,V,W$). 
In this coordinate system
$X$ points from the Sun in direction of the galactic centre 
with its origin in the galactic centre, 
$Y$ points into the direction of the galactic rotation at the position of the sun, and 
$Z$ toward the north galactic pole. 
The same applies to the corresponding velocities $U$, $V$, $W$. 

The orbits are calculated using the model for the gravitational potential
of our Milky Way by Allen \& Santillan (1991) which was developed 
to be used in an orbit calculating program (Odenkirchen \& Brosche 1992).
This model has been extensively used in the studies of 
de Boer \ea\ (1997a), Geffert (1998) and Scholz \ea\ (1996). 
There are several other models available which yield similar results 
as long as the orbits do not extend to extreme distances from the 
galactic centre (Dauphole et al. 1996).
The model of Allen \& Santillan (1991) is based on 
$\Theta_{\rm LSR} = 220$ \kms\ and $R_{\rm LSR} = 8.5$ kpc.
The values for the peculiar velocity of the Sun used in the calculations in 
this paper are $U_{{\rm pec},\odot}=10$ \kms, $V_{{\rm pec},\odot}=15$ \kms,
$W_{{\rm pec},\odot}=8$ \kms.

\begin{table*}
\caption[]{Mean orbital parameters $\Theta$, $ecc$ and $nze$ for various subsamples of HB stars}
\begin{tabular}{llcrrrrrr}
\hline
Types  &    subsample            & number   & $nze$ & $\sigma_{nze}$ & $\Theta$ & $\sigma_{\Theta}$ & $ecc$ &  $\sigma_{ecc}$ \\
       &                         & of stars &       &                & [\kms]   & [\kms]            &       &                 \\
\hline
HBA    &      this paper         &   14     & 1.10  &     1.15       &   +17   &    52           & 0.74  &      0.16       \\ 
HBB    & this paper \& Schmidt (1996) &    6     & 0.24  &     0.15  &  +151   &    55           & 0.41  &      0.27       \\ 
sdB/O  &      this paper         &    5     & 0.04  &     0.05       &  +218   &    10           & 0.12  &      0.08       \\ 
sdB    & this paper \& de Boer (1997b) &   44     & 0.25  &   0.17   &  +198   &    50           & 0.15  &      0.11       \\ 
RR Lyr &      all                &   32     & 0.86  &     1.50       &   +80   &   114           & 0.59  &      0.33       \\ 
RR Lyr & [Fe/H] $>-0.9$        &    7     & 0.08  &     0.05       &  +218   &    37           & 0.19  &      0.13       \\ 
RR Lyr &  $-0.9>$[Fe/H]$>-1.3$ &    7     & 0.88  &     0.57       &  +43    &   109           & 0.64  &      0.38       \\ 
RR Lyr &  $-1.3>$[Fe/H]$>-1.6$ &   10     & 1.54  &     2.43       &  +32    &    74           & 0.68  &      0.40       \\ 
RR Lyr &    $-1.6<$[Fe/H]      &    8     & 0.68  &     0.47       &  +51    &   107           & 0.65  &      0.30       \\ 
\hline
\end{tabular}\\
\noindent
\end{table*}

To determine the parameters $z_{\rm max}$, 
the maximum height reached above the galactic plane and
$R_a$ and $R_p$, the apo- and perigalactic distances, 
we calculated the orbits over 10 Gyr. 
This for certain does not give true orbits as the orbits are 
probably altered in time by heating processes.
However this long timespan allows to better show the area
the orbit can occupy in the meridional plane (see Fig. 2).

As in de Boer et al. (1997a), we also calculated the eccentricity $ecc$ 
of the orbit, given by
\begin{equation}
ecc=\frac{R_{\rm a}-R_{\rm p}}{R_{\rm a}+R_{\rm p}}
\end{equation}
 and the normalised $z$-extent, $nze$, given by
\begin{equation}
nze=\frac{z_{\rm max}}{\varpi(z_{\rm max})}.
\end{equation}
The parameter $nze$ is more relevant than $z_{\rm max}$, 
since it accounts for the effect of diminished gravitational potential 
at larger galactocentric distance $\varpi$.

To assign a star to a population often the $U,V,W$-velocities 
and their dispersions are used, as well as the orbital velocity $\Theta$. 
For stars near the Sun (small $Y$), the $V$ velocity 
is nearly the same as $\Theta$. 
However, for stars further away from the Sun's azimuth,
$\Theta$ becomes a linear combination of $U$ and $V$. 
Therefore $\Theta$ should be preferred.
In order to make comparisons with results from other studies, 
we use both $U,V,W$ and $\Theta$.

We calculated the errors of the velocity components and the orbital velocity
using Monte Carlo simulations of Gaussian distributions to vary the
input parameters within their errors as described by Odenkirchen (1991). This is neccesary
rather than just calculating errors using Gauss error propagation laws because
the parameters are significantly correlated. For the error calculation we used 
the software of Odenkirchen (priv. comm.). The proper motion errors were 
taken from the Hipparcos catalogue. The errors of the distances were calculated from the
error in absolute magnitude as derived in Sect. 2.3. We took the errors of the radial velocities
as published in the respective articles. For those radial velocities of Wilson (1953) and Evans (1967)
having quality mark ``e'', meaning the error is larger than 10 \kms, we used 15 \kms\ as error.
This is justified as can be seen by comparison of these values with those of other studies.
Generally the error in the velocity components is less than 10 \kms. Only a few stars have somewhat larger
errors, the largest error in $\Theta$ being 12 \kms. For the HBA/B stars the typical value of $\Delta\Theta$
is about 7 \kms, for the on average closer sdB/O stars $\Delta\Theta$ is 1 to 2 \kms.

We estimated errors for $nze$, $ecc$, $R_{\rm a}$ and $R_{\rm p}$ 
because they have not been used individually in the interpretation.
Moreover the larger values of $nze$ are very sensitive to small variations in the shape of the orbit. This
especially applies to stars having chaotic orbits. 
Variations in the input distance modulus showed that the resulting variations in all of these quantities except
$nze$ are relatively small in most cases. For a discussion of overall effects on a sample see de Boer \ea\ (1997a).

\subsection{Morphology of the orbits}
The orbits of the HBA/B stars show a large variety of shapes.
Nearly all of the cooler HBA stars have a small perigalactic distance 
($R_{\rm p}\le$ 3 kpc) and the most extreme case, 
HD~86986, reaches a perigalactic distance of only 0.4 kpc. 
The single exception is HD 117880, which has a $R_{\rm p}$ of nearly 4 kpc.

Four stars have truly chaotic orbits, 
the rest has boxy type orbits, but some of these show signs of chaotic
behaviour as well. HD 79813 has an orbit staying very close to the disc, while HD 117880
orbits nearly perpendicular to the galactic plane. 
On the whole about half of our stars have orbits which are chaotic or show 
signs of that. 
This agrees quite well with the results of Schuster \& Allen (1997) 
who analysed a sample of local halo subdwarfs.

Most of the stars have apogalactic distances of $\simeq$ 8 to 11 kpc, 
just one star (HD~86986) goes well beyond.
The reason for this clumping in $R_{\rm a}$ is not physical but due to 
selection effects. 
Stars with $R_{\rm a}\le 7.5$ kpc never venture into the observable zone 
(at least observable by Hipparcos).
On the other hand the probability of finding the stars is greatest 
when they are near their major turning point, $R_{\rm p}$. 
So it is clear that the mean $R_{\rm a}$, as well as to a lesser extent the eccentricity,
are affected by selection effects.

Stars belonging to the thin disk would have orbits with 
very small eccentricities and $nze$ values
(solar values: $ecc$=~0.09, $nze$=~0.001, see de Boer \ea\ 1997a),
while thick disk stars would have larger values on average. 
Halo stars have generally orbits with large eccentricities 
while their $nze$ show a large range.

The eccentricities of the HBA star orbits are very large, 
ranging from 0.5 to nearly 1.0, 
the values for $nze$ vary by a huge amount, 
from 0.04 (HD~78913) to 5 (HD~117880).
The stars BD~+36 2242 and Feige 86 are exceptions, 
their values for both parameters are more appropriate for disk objects.
We note that these two stars are the hottest of the HBA/B sample. 
The kinematics of the four HBB stars from Schmidt (1996) show overall
behaviour similar to that of BD~+36 2242 and Feige 86 (Fig. 2). 
All of these are hotter than 11000 K, the $T_{\rm eff}$ of BD~+36~2242. 

The star HD 117880 features an orbit somewhat dissimilar from the others. 
While its $nze$ is very high, its eccentricity is by far the lowest 
of the sample of HBA stars. 

\begin{figure*}
\def\epsfsize#1#2{1.0\hsize}
\centerline{\epsffile{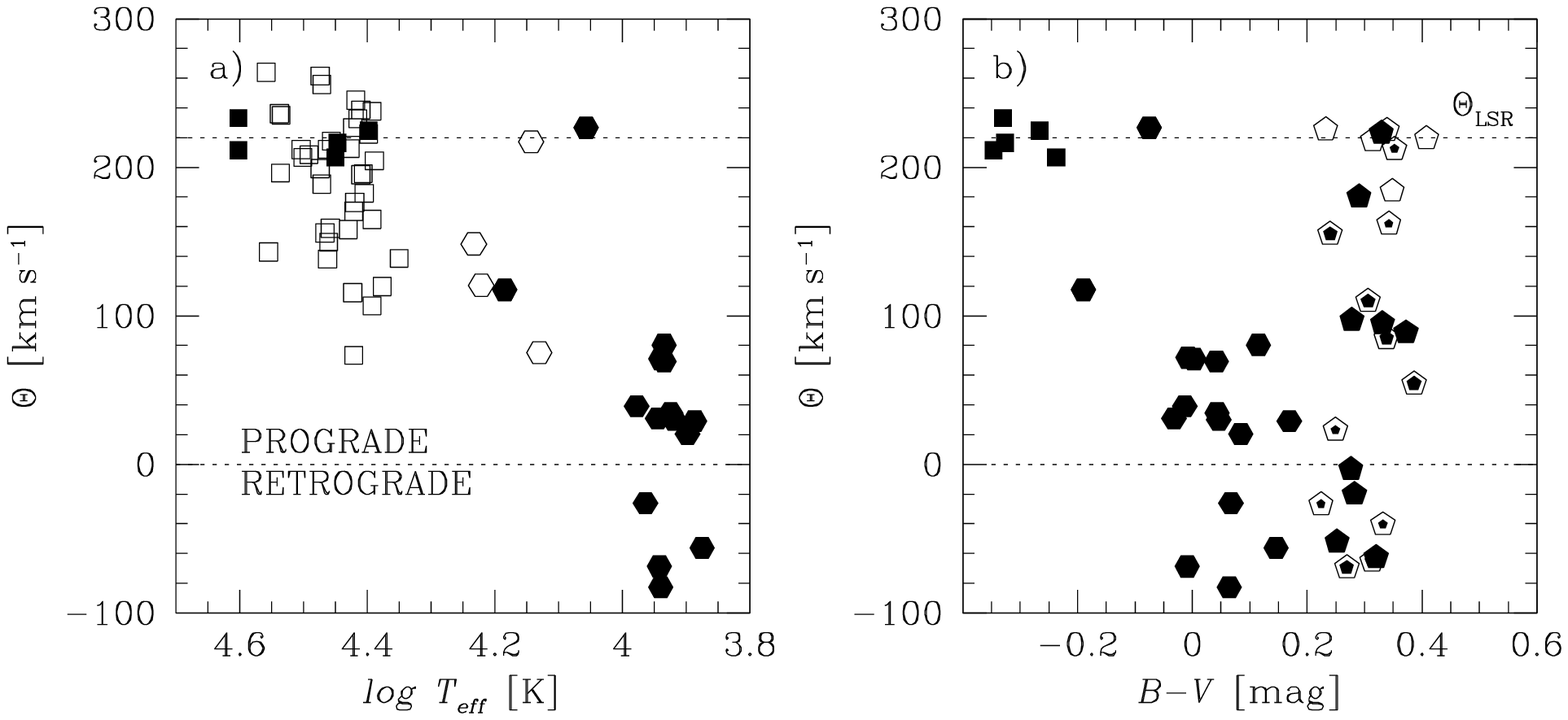}}
\caption[]{ {\bf a-b}. 
Kinematical trend of stars along the field horizontal branch 
characterized by the orbital velocity ($\Theta$) 
as plotted against effective temperature \teff\ and $B-V$.\\
Panel {\bf a:} $\Theta$ versus $T_{\rm eff}$. 
Filled symbols show the stars with Hipparcos data (Table 3), 
open symbols are sdB stars of de Boer \ea\ (1997a)
and HBB stars of Schmidt (1996). 
The symbol shapes represent: hexagons HBA/B stars; squares sdB/O stars.\\
Panel {\bf b:} $\Theta$ versus $B-V$, highlighting the cooler part 
of the HB and including RR Lyrae stars (see Sect.~4).
The RR-Lyrae stars are plotted with pentagons subdivided according to 
[Fe/H]$<-$1.6~dex (full), $-1.6$~dex~$\le$[Fe/H]$<~-$1.3~dex,
$-1.3$~dex~$\le$[Fe/H]$<~-$0.9~dex
and [Fe/H]$\ge~-$0.9~dex (open symbols)
}
\end{figure*}

\subsection{Velocity components and dispersions}
The HBA stars (\teff\ $\le$ 10,000 K) have a mean orbital velocity of 
$\Theta$ = 17 \kms, lagging about 200 \kms\ behind the local standard of rest.
However, the velocity dispersions are large: 
102, 53 and 95 \kms\ in $U$, $V$, $W$ respectively. 
This shows that there are many stars with a non disk-like 
kinematical behaviour in the sample of HBA/B stars. 
They therefore belong to the galactic halo population rather than to the disk.
    
The orbital velocities of the HBA stars in the sample do not have a 
Gaussian distribution, as one might have expected.
Instead, they seem to have a somewhat flatter distribution (see Fig. 4).
 About 75\% of our stars have prograde velocities, 
four stars have retrograde orbits. 
However the exact distribution cannot be studied reliably
due to the limited number of stars at disposal.

Both the analysis of the kinematic properties and the shapes of the orbits 
imply that the HBA/B stars mostly are members of the galactic halo population. 
However, there seems to be a difference in kinematics and 
hence population membership between the cooler and the hotter stars. 
Stars cooler than about 10,000 K have low orbital velocities 
and a large spread in $nze$.
In contrast to this are the hotter stars whose kinematics and orbits 
are consistent with those of disk objects. 
The HBB stars of Schmidt (1996) which are all hotter than 10,000 K 
behave like sdB stars, too. 

\subsection{Kinematics of sdB/O stars}
The sample of sdB/O stars show classical disk behaviour: 
Their mean orbital velocity is $\Theta=219$ \kms, 
meaning a negligible asymmetric drift. 
The $V$ velocity dispersion (which is also the dispersion in $\Theta$, 
because the stars are in the solar vicinity)
 is relatively small, similar to that of old thin disk orbits, 
while the dispersion in $U$ is much larger, fitting to thick disk values. 
The dispersion $\sigma_W$ is somewhere in between. 
These values are quite similar to those of the sdB star sample of de Boer \ea\ (1997a).
Until now no population of {\it field} sdB stars with halo kinematics 
has been found. 
Yet, hot subdwarfs of the horizontal branches of halo globular 
clusters are, of course, well known (see e.g. Moehler \ea\ 1997).
 
\subsection{Trend of kinematics along the HB?}
Given the results above there seems to be a trend in the kinematics 
of star types along the blue part of the horizontal branch (see Fig. 3).
The sdB/O stars have disklike orbits. 
The same probably applies to the HBB stars hotter than about 10,000 K, 
though the statistics are rather poor for this part of the HB.
In contrast to that stand the cooler HBA stars which have much 
smaller orbital velocities, large orbital eccentricities and 
large ranges of $nze$, 
thus showing a behaviour fitting more to halo than to disk objects. 

This result suggests to analyse the kinematics of the adjoining
cooler stars of the HB, the RR Lyraes.

\begin{figure}
\def\epsfsize#1#2{1.0\hsize}
\centerline{\epsffile{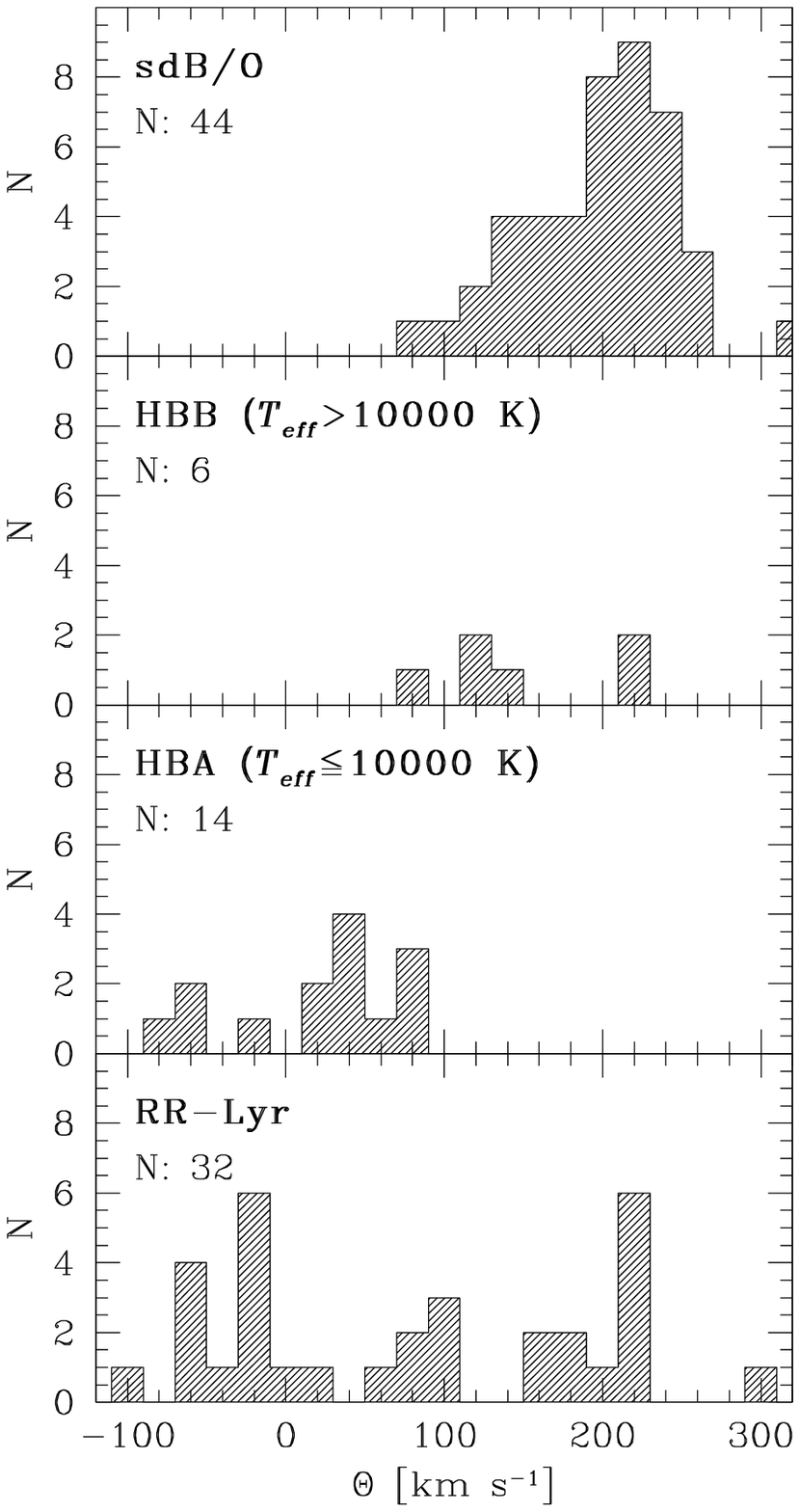}}
\caption[]{Histogram showing the distribution of orbital velocities of the investigated stars.
The binsize is 20 \kms.}
\end{figure}

\section{RR Lyrae stars}

\subsection{A sample of RR Lyrae stars from the literature}
Recently, Martin \& Morrison (1998) carried out an investigation 
of the kinematics of RR Lyrae stars
which is mainly based on the study of Layden (1994).
For our analysis we will use only those stars having Hipparcos data.
Six Hipparcos stars were excluded because they have a proper motion error 
larger than 5 mas/yr.

The RR Lyrae stars present the observational difficulty in that they are variables with both $V$ and
$B-V$ changing continously.
For most of the sample we were able to take the mean magnitudes from Layden (1994).
For the remaining stars we
derived the intensity-mean magnitudes with help of the formula given by Fitch \ea\ (1966) and revised
by Barnes \& Hawley (1986) which is the same method as used by Layden (1994), using the photometric data of Bookmeyer \ea\
(1977). The Layden photometry was dereddened using the Burstein \& Heiles (1982) reddening maps.

For later steps in this study it is necessary to know the mean $B-V$ 
of the RR Lyrae stars. 
As the colour curves of the stars are quite similar to the brightness curves, 
with the star being bluest when it is near maximum brightness, 
we took the same formula as we used to calculate the mean magnitude. 
This is not entirely correct but gives $B-V$ close to the actual one.
For six stars we did not have the appropiate light curve data, so we could not
determine the mean $B-V$ for them. Therefore only 26 RR Lyraes are shown in Fig. 3.

As the RR-Lyrae stars are in most cases fainter and therefore farther away 
than our HBA/B stars they have a rather large $\Delta_{\pi}/\pi$. 
For this reason we used the absolute magnitude derived in Sect. 2.3 
to calculate the distances for these stars. We thus have ignored the effects 
of metallicity  on $M_V$ for individual stars. Also possible 
evolutionary effects on $M_V$ (see Clement \& Shelton 1999) have been ignored,
an aspect Groenewegen \& Salaris (1999) did not consider in their determination
of the RR Lyrae $M_V$ either.
Since we study the orbits of the RR Lyrae as a sample these limitations will
 not affect our conclusions.

For most RR Lyrae stars we took the radial velocities from the sources 
mentioned in Sect. 2.5, 
supplemented by radial velocities from Layden (1994).

The metallicities of the RR-Lyrae stars were taken from Layden (1994) as far as possible. A few values come from
Layden et al. (1996) and Preston (1959).

\subsection{RR-Lyrae kinematics}
We calculated the orbits for the RR Lyrae stars in the same manner 
as for the HBA/B and sdB stars.
The RR-Lyrae stars show a spread in kinematical behaviour wider than that 
of the HBA/B stars. 
Many stars have orbits similar to those of the HBA stars,
others show disklike orbits with orbital velocities in the 
vicinity of 200 \kms. 
Of the halo RR-Lyrae stars many have perigalactic distances 
smaller than 1 kpc, as we also found for the HBA stars.
The RR-Lyrae stars have orbital velocities typically 
spanning the entire range found for disk and halo stars (see Fig. 3). 
Three members of our sample of RR Lyr stars have orbits shaped somewhat 
different from those of the rest of halo orbits,
looking similar to that of HD 117880.

In Fig. 3 we have sorted the RR Lyr stars according to their metallicity 
using different plot symbols.
The stars with an [Fe/H]$> -0.9$ dex
have high $\Theta$ like disk stars. The stars with lower metallicities are more evenly distributed in $\Theta$.
There are several stars with disk-like kinematics with a very low metallicity as low as [Fe/H]$< -2.0$ dex (see Table 4).

\section{Selection effects}

The study of the spatial distribution of HB stars involves, 
unfortunately, several selection effects. 
The general aspects have been reviewed by Majewski (1993) and will 
not be repeated in detail here. 
Yet, for each stellar type discussed in this paper a few 
comments are in place. 

HBA stars have in most cases been identified from photometry, 
notably because of a larger than normal Balmer jump. 
This larger jump is mostly due to lower metallicity of the 
stellar atmosphere. 
If the atmospheric metallicity is identical to the original one, 
then the criterion favours intrinsically metal poor stars, 
which are presumably the older ones. 
However, also stars starting with a little more mass than the Sun 
and thus of solar composition will become HB stars and, 
when as old as the Sun, by now are solar metallicity HB stars. 
If they were of HBA type, they would not have been recognized in 
photometry of the Balmer jump. 
Such stars would be underrepresented in our sample. 
The HBA stars considered here come from all galactic latitudes, 
so that selection effects due to galactic latitude are not to be expected. 
However stars which have orbits going far away from the disk 
are always underrepresented, 
as their fraction of time near the disk (and hence being observable) 
is much smaller than for those which do not go far from the disk.

RR Lyraes, being variables, are not prone to such selection effects. 
Most of them are identified solely by their variability. 
Metallicity or high velocity are generally not used as criteria for the 
identification for RR Lyrae stars. 
For a discussion of selection effects due to galactic latitude we refer 
to Martin \& Morrison (1998), as our sample is a subsample of theirs. 

The sdB/O stars were identified in surveys for quasars, e.g. 
the PG catalogue (Green \ea\ 1986) or Hamburger Quasar Survey 
(Hagen \ea  1995). 
This means their blue colour is the criterion, rather than proper motion, 
radial velocity or metallicity. 
Therefore we do not expect a selection bias towards metal poor halo stars. 
Moreover, de Boer \ea\ (1997a) showed that the sdB/O stars observed
now near the Sun come from widely differing locations in the Milky Way.
As these catalogues only map objects which are somewhat away from 
the galactic plane, they miss the majority of stars with solar type orbits. 
sdO stars may be confused with pAGB stars descending down
the HRD towards the white dwarf regime.

The HBB stars of Schmidt (1996) are also taken from the PG catalogue, 
so that there should not be noticeable selection effects, either. 
However HBB stars and main sequence stars have similar physical properties 
such as \lgg, so that there may be confusion with the latter. 
Apart from this the selection effects mentioned for the sdB/O stars 
apply to the HBB stars, too. 
 
Finally, some words concerning the distribution of distances of the 
different samples are in place. Generally, if one deals with stars having different
absolute magnitudes, as in our case when the sdBs are several magnitudes
fainter than the HBAs, one gets samples with different mean distances. The intrinsically fainter stars are on average much nearer
than the brighter stars, if the two groups have similar apparent magnitudes. 
This means that the spatial regions sampled differ depending on 
the absolute magnitude of the stars.
This would imply that the sdB sample is biased towards disk stars 
as we do not sample them far enough from the galactic plane where there  
may be a higher concentration of halo stars than further in. 
This is however not the case. As we include some of the results of 
de Boer et al. (1997a) which come from a completely different source, 
namely mostly from the PG-catalogue (Green et al. 1986) dealing with 
significantly fainter stars, the PG stars actually have on average 
larger distances than any of our HBA stars. 
For this reason we do not expect that the difference in kinematics 
arises from the distribution of the distances in the samples.

\section{Discussion: trends and population membership}
\subsection{Overall trends}
As shown in Figs. 3 and 4
the kinematics of the stars of horizontal branch type appears 
to have a trend along the HB indeed. 

The {\it sdB stars} have in general rather disk-like orbits 
and kinematical properties. 
The ones analysed here (Table 4) show the same behaviour as those from 
the large sample of sdB stars investigated previously (de Boer \ea 1997a).

The {\it HB stars}, the prime goal of our investigation, 
span a wide range in orbit parameters but when this group 
is split in HBB and HBA stars a cut is present. 

The (hotter) {\it HBB stars} behave rather like the sdBs 
with orbits of disk-like characteristics.  
However, such stars are difficult to recognize and our sample is small. 
A much larger sample may show a larger variation in kinematics.

The {\it HBA stars} have mostly halo orbits (mean $\Theta \simeq 17$~\kms).
This is very similar to the value at which most other studies concerning 
metal poor stars in the solar neighbourhood arrive 
(see Table 2 of Kinman 1995). 
However, the known sample may be observationally skewed toward 
stars with low atmospheric metallicity (large Balmer jump).
 
The {\it RR-Lyrae stars} have orbits spanning a large range
in orbital parameters, too.
However, a trend seems to be present with metallicity.
The metal poor stars have halo orbits similar to those of the HBA stars with
rather low orbital velocities of less than 100 \kms,
and large $ecc$ and $nze$.
The metal rich stars on the other hand have rather  
disk-like kinematical characteristics. 
A similar distribution of metallicities and orbital velocities 
was also found in the studies of Chen (1999) and Martin \& Morrison (1998).

 Although there are a few RR Lyraes having high orbital velocities 
($\Theta \ge 160$ \kms) and clearly disk-like orbits 
(some of which are very metal poor), 
HBA stars with such characteristics are not found in our sample. 
On the other hand no RR Lyraes with [Fe/H]$> -0.9$ dex with halo-like orbits 
or kinematics are present. 
This means that a high metallicity for a RR Lyr star is a good indicator 
that it is a disk star. 
However, a low metallicity does not mean that a star neccessarily belongs 
to the halo. 

For an overview of literature data on values for $\Theta$ 
(or asymmetric drift) for various star groups 
we refer to Fig.\,3 in the review of Gilmore \ea\ (1989).

\subsection{Discussion}

Since the sdB stars (and possibly the HBB stars) have disk-like orbits, 
these stars must be part of a relatively younger, more metal rich group among the HB stars. 
Majewski (1993) uses the expression `intermediate Population II', 
other authors use the words `thick' or `extended disk'. 
In addition to the disk-nature of their orbits, 
the vertical distribution is consistent with a scale height 
of the order of 1 kpc (Villeneuve \ea\ 1995, de Boer \ea\ 1997a). 
Since the amount of metals in their atmospheres may have been altered by 
diffusion it is not possible to estimate the true age from the metallicity. 

The HBA stars have really halo orbits. 
This must mean they belong to a very old population. 
Their atmospheric metal content is low indeed, 
the determinations showing a large scatter per star and 
from star to star ranging between $-1$ and $-2$ dex. 
However, metal rich HBA stars 
which are known to exist in star clusters (see Peterson \& Green 1998), 
would likely be underrepresented in the sample. 

If the halo contains mostly old stars, like globular cluster stars, 
then the resulting halo HB stars should occupy the HB in ranges 
related with metallicity as with the globular clusters (see Renzini 1983). 
The very metal poor ones ([$M/H$] $\simeq -2$ dex) would be HB stars of 
HBB and HBA nature as well as RR Lyrae, 
the ones of intermediate metallicity ([$M/H$] $\simeq -1.5$ dex) 
would be very blue down to sdB like,
and the metal rich ones ([$M/H$] $\simeq -1$ dex) 
would be RHB stars, perhaps including some RR Lyrae. 
This behaviour may also explain the existence of the two Oosterhof groups 
(see van Albada \& Baker 1973 or Lee \ea\ 1990) of RR Lyrae, 
since only the very metal poor and the relatively metal rich 
globular clusters contain RR Lyrae. Evolutionary changes of the HB stars
may also affect the location on the HB (Sweigart 1987, Clement \& Shelton 1999). 

However, sdB stars with halo kinematics have not been found 
(de Boer \ea\ 1997a). Instead, they have only disk orbits. 
This must mean that the stars which originally formed in the halo had
an initial mass, a metallicity and a red giant mass loss such that RR Lyrae and HBA stars 
were the end product, and not sdB stars.

As for the RR Lyrae stars, they show a wide range in kinematic behaviour, 
more or less in line with the atmospheric metal content. 
The actual metallicity did not bias the identification of these stars, 
since they are selected based on variability. 
One tends to divide the RR Lyrae sample into metal poor and 
metal rich RR Lyrae (see Layden 1994). 
Here we recall that in the HB stars the contents 
of heavier elements in their atmospheres may be altered (see Sect. 1).
The RR Lyrae stars with the continuous upheaval 
of the pulsation may stimulate mixing 
so that their atmospheres probably show the true metallicity. 
Thus, for RR Lyraes the metallicity may be used as a general population tracer. 
The observed range of metallicities would mean 
that there are old as well as younger RR Lyraes. 
  
Old RR Lyrae must be very metal poor and should have halo orbits. 
The majority of the RR Lyrae included in our analysis fit these parameters. 
There are, however, a substantial number of RR Lyr stars in our sample 
with disk-like kinematics but low metallicities, 
in several cases as low as $-2$ dex. 
The origin of this group of stars, dubbed the `metal weak thick disk', 
is still unknown (see Martin \& Morrison 1998 for a discussion). 

Young (or younger) RR Lyrae should be relatively metal rich 
and have disk orbits. The investigated sample contains such stars. 
These objects should have an age, main-sequence mass, metallicity 
and RGB mass loss such that RR Lyrae emerge, 
i.e. HB stars with a thicker hydrogen shell. 
They are, being relatively metal rich, 
also of slightly different $M_V$ than the metal poor and old ones. 
In fact, they are fainter and their distances should be based 
on the appropriate brightness-metallicity relation. 
The dependence is, however, feeble and amounts to 
just 0.1 mag for 0.5 dex. 
We tested how serious ignoring this effect is on the derived orbits 
by reducing the RR Lyr star distances by 10 \%. 
It does not lead to a change of significance in the histogram of Fig.\,4.

\subsection{Summary}

Our orbit studies allow to see a trend in the kinematics 
of the field HB stars along the horizontal branch. 
This appears to give us access to the structure of 
the Milky Way and its halo as well as information about possible formation scenarios. 
The trends related with age and history could only be found using the kinematics, 
since it has become clear that the atmospheric metallicity in HB-like stars 
has no relation to the one of the main sequence progenitor. 
The location of the stars on the HB must be a complicated function of 
age, main-sequence mass, initial metallicity, and mass loss on the RGB. 
For the HB-like stars of today indications for the age can be determined 
from the present kinematic parameters. 
Only detailed models for metallicity dependent stellar evolution 
from main sequence through the RG phase with mass loss should, 
in comparison with the observables of horizontal branch stars, 
eventually be able to retrieve the true origin of the HB stars.

\acknowledgements{
We thank Oliver Cordes for supplying the values of \lgg\ and \teff\ 
for two stars. 
We are very grateful to Michael Odenkirchen
who supplied us the orbit calculating software.
Furthermore we thank Michael Geffert for enlightening discussions, 
Wilhelm Seggewiss and J\"org Sanner for carefully and critically reading the manuscript. 
This research project was supported in part by the 
Deutsche Forschungs Gemeinschaft (DFG) under grant Bo 779/21. 
For our research we made with pleasure use of the SIMBAD in Strasbourg.
}
%

\end{document}